\documentclass[utf8]{frontiersFPHY} 

\setcitestyle{square} 
\usepackage{url,hyperref,lineno,microtype,subcaption}
\usepackage[onehalfspacing]{setspace}
\usepackage{bm}
\usepackage{amsfonts}%
\usepackage{amssymb}
\usepackage{amsmath}%
\usepackage{color}%

\def\e{{\epsilon}}
\def\k{{ {\bm k} }}
\def\p{{ {\bm p} }}
\def\q{{ {\bm q} }}
\def\Q{{ {\bm Q} }}

\def\0{{ {\bm 0} }}
\def\w{{\omega}}
\def\a{{\alpha}}

\def\keyFont{\fontsize{8}{11}\helveticabold }
\def\firstAuthorLast{Onari {et~al.}} 
\def\Authors{Seiichiro Onari\,$^{1,*}$ and Hiroshi Kontani\,$^{1}$}
\def\Address{$^{1}$Department of Physics, Nagoya University,
Furo-cho, Nagoya 464-8602, Japan.}
\def\corrAuthor{Seiichiro Onari}

\def\corrEmail{onari@s.phys.nagoya-u.ac.jp}

\begin{document}
\onecolumn
\firstpage{1}

\title[Exotic orders in Fe-based superconductors]{Diverse
Exotic Orders and Fermiology in Fe-based Superconductors: A Unified
Mechanism for $B_{1g}/B_{2g}$ Nematicity in FeSe/(Cs,Rb)Fe$_2$As$_2$
 and Smectic Order in BaFe$_2$As$_2$} 

\author[\firstAuthorLast ]{\Authors} 
\address{} 
\correspondance{} 

\extraAuth{}

\maketitle

\begin{abstract}
\section{}
\color{black}A rich variety of nematic/smectic orders in Fe-based superconductors is an
important unsolved problem in strongly correlated electron systems. \color{black}
A unified understanding of these orders has been investigated for
the last decade. In this article, we explain the $B_{1g}$ symmetry
 nematic transition in FeSe$_{1-x}$Te$_x$, the $B_{2g}$ symmetry nematicity in AFe$_2$As$_2$ (A=Cs, Rb),
 and the smectic state in BaFe$_2$As$_2$ based on the same framework.
\color{black}We investigate the quantum interference mechanism between spin
fluctuations by developing the density wave equation. The observed rich
variety of nematic/smectic orders is naturally understood in this
mechanism. The nematic/smectic orders depend on the characteristic
shape and topology of the Fermi surface (FS) of each compound. \color{black}
(i) In FeSe$_{1-x}$Te$_x$ $(n_d=6.0)$, each FS is very small and the
$d_{xy}$-orbital hole pocket is below the Fermi level. \color{black}In this case, the small spin
fluctuations on three $d_{xz}$, $d_{yz}$, and $d_{xy}$ orbitals
cooperatively lead to the $B_{1g}$ nematic $(\q=\bm{0})$ order without
magnetization. \color{black} The
experimental Lifshitz transition below the nematic transition
temperature $(T_S)$ is naturally reproduced.
(ii) In BaFe$_2$As$_2$ $(n_d=6.0)$, the $d_{xy}$-orbital hole pocket emerges around M point, and
each FS is relatively large. The strong spin fluctuations due to
the $d_{xy}$-orbital nesting give rise to the $B_{1g}$ nematic
$(\q=\bm{0})$ order and the smectic $[\q=(0,\pi)]$ order, and the latter
transition temperature ($T^*\sim 170$K) exceeds the former one ($T_S\sim140$K).
(iii) In heavily
hole-doped AFe$_2$As$_2$ $(n_d=5.5)$, the large $d_{xy}$-orbital hole
pocket and the four
tiny Dirac pockets appear due to the hole-doping.
The $B_{2g}$ nematic bond order emerges on the $d_{xy}$-orbital
hole pocket due to the same interference mechanism. 
\color{black}The present paramagnon interference mechanism provides a
unified explanation of why the variety of nematic/smectic orders in Fe-based
superconductors is so rich, based on the well-established fermiology of
Fe-based superconductors. \color{black}

\tiny
 \keyFont{ \section{Keywords:} Nematic state, Smectic state, Orbital order, Bond order, Quantum
 critical point, Fe-based superconductors} 
\end{abstract}

\section{Introduction}
The emergence of electron nematic ($\q=\bm{0}$) state is one of the most important
unsolved problems in Fe-based superconductors \cite{Hosono-review}. In
LaFeAsO and Ba122 compounds, the antiferro (AF) magnetic state appears at the N\'eel temperature $T_N$, which is lower than
the nematic transition temperature $T_S$.
Since the superconducting
phase with high transition temperature $(T_c)$ appears near the nematic
phase and the AF magnetic phase, it is expected that the nematic
fluctuations and the spin fluctuations are
related to the mechanism of the high-$T_c$ superconductivity.
However, the questions appear before discussing the
superconductivity: (i)What is the order parameter of the nematic state?
(ii)What is the driving force of the nematic state? (iii)Why do the diverse
nematic states emerge in various compounds.

It is known that the nematic order cannot be derived from the mean-field
theory since the spin-channel order always dominates over the nematic
order unless unphysical parameters (such as the negative Hund's coupling) are assumed.
Previously, in order to explain the nematic state
\cite{Hirschfeld-review}, the vestigial order (spin
nematic) scenario
\cite{Fernandes,Fernandes-122,DHLee,QSi,Valenti,Fang,Fernandes-review}
and the orbital
order scenario \cite{Kruger,PP,WKu,Kontani-PRL,Onari-SCVC,Onari-Hdope,Onari-form,Yamakawa-PRX,Yamakawa-FeSe-super,Onari-B2g,Onari-122,JP,Fanfarillo}
have been proposed. 

To investigate the nematic state, the FeSe
family is an ideal
platform since
the AF magnetic state is absent
\cite{FeSe,Bohmer-FeSe-review,Andersen-FeSe-review,Shibauchi-FeSe-review}.
This family is also attractive from the aspect of superconductivity since the
highest $T_c\gtrsim65$K in Fe-based
superconductors has been reported in electron-doped FeSe \cite{eFeSe1,eFeSe3,eFeSe4,eFeSe5,eFeSe6}. 
In FeSe, the orbital polarization between $d_{xz}$ and
$d_{yz}$ orbitals in the nematic state has been observed by the angle-resolved-photoemission
spectroscopy (ARPES) \cite{FeSe-ARPES-Shimojima,FeSe-ARPES-Takahashi,FeSe-ARPES-Suzuki,FeSe-ARPES-Shen,FeSe-ARPES-Shen2}.
To be more precise, the orbital polarization energy $E_{xz}-E_{yz}$ has $\k$ dependence and changes
the sign between $\Gamma$ point and X(Y) point. 
This sign reversal orbital polarization has been explained by the orbital
order scenario \cite{Onari-form,Yamakawa-PRX,Onari-B2g,Onari-122} based on the paramagnon interference
mechanism and by the renormalization group (RG)
theory \cite{Chubukov-RG1,Chubukov-RG}. 
In both theories, the vertex correction (VC) of the Coulomb interaction,
which corresponds to the higher-order many-body effect, plays the essential role. Since the AF magnetic
correlation is weak in FeSe,
it is difficult to explain the nematic state by the
vestigial order (spin
nematic) scenario.
Based on the paramagnon interference mechanism, the
$B_{1g}$ nematic orders in LaFeAsO and FeSe
\cite{Onari-SCVC,Onari-Hdope,Onari-form,Yamakawa-PRX,Yamakawa-FeSe-super} and the nematic orders
in cuprate superconductors \cite{Kawaguchi,Yamakawa-Cu,Tsuchiizu-Cu} and
magic-angle twisted bilayer graphene \cite{Onari-TBG} have been
explained as the orbital/bond orders. CDW orders in the transition metal
dichalcogenide \cite{Hirata} and kagome metal \cite{Tazai-kagome} have also been
explained by the paramagnon interference mechanism.

\begin{figure}[!htb]
\includegraphics[width=.9\linewidth]{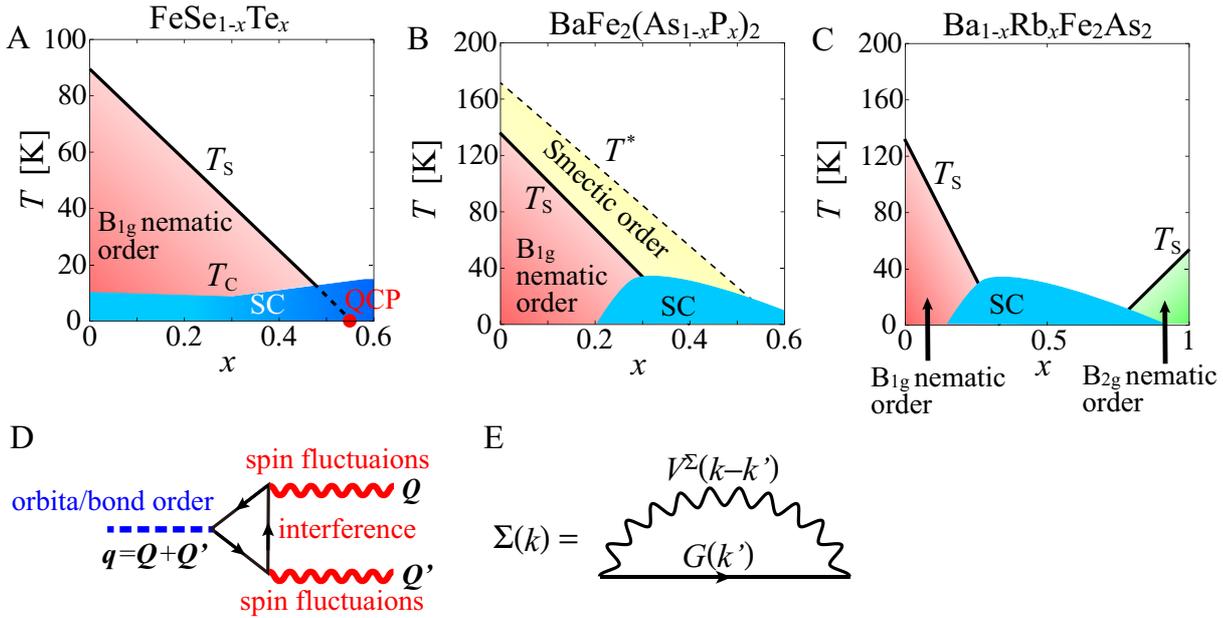}
\caption{
(A) Schematic $x$-$T$ phase diagram of FeSe$_{1-x}$Te$_{x}$, where $T_S$
 decreases with $x$, and $T_c$ becomes maximum near the nematic QCP.
(B) Schematic $x$-$T$ phase diagram of BaFe$_2$(As$_{1-x}$P$_{x})_2$,
 where the tiny nematicity appears for $T_S<T<T^*$. We explain that
 ``tiny nematicity'' above $T_S$
 originates from smectic bond order in later section.
(C) Schematic $x$-$T$ phase diagram of Ba$_{1-x}$Rb$_{x}$Fe$_2$As$_2$.
$B_{2g}$ nematic order appears for heavily hole doped region
 $x > 0.5$.
(D) Feynman diagram of the paramagnon interference mechanism for the
 orbital/bond order.
(E) Feynman diagram of the self-energy $\Sigma(k)$.
}
\label{fig:interference}
\end{figure}

\color{black}The rich variety of the nematicity in the FeSe family remains a significant
open problem. \color{black} In FeSe$_{1-x}$S$_x$, $T_S$ disappears at
$x\sim0.17$, where the emergence of the nematic
quantum critical point (QCP) has been suggested by experiments
\cite{QCP-FeSeS1,QCP-FeSeS4,QCP-FeSeS2,QCP-FeSeS3}. 
Recently, whole $x$ dependent phase diagram for FeSe$_{1-x}$Te$_x$ ($x\lesssim
0.6$) has
reported \cite{FeSeTe,FeSeTe2,FeSeTe3}. In the phase diagram shown in
Fig. \ref{fig:interference}A, $T_S$ decreases with Te doping $x$, and $T_S$ disappears at
$x\sim0.5$. 
$T_c$ becomes maximum $\sim 15$K at $x\sim 0.6$,
which indicates that the nematic fluctuations enlarge superconducting pairing
interaction near the nematic QCP. \color{black}Thus, it is essential to clarify the mechanism of $x$ dependence
of $T_S$ to understand the mechanism of
superconductivity in the FeSe family. \color{black}

In addition, a significant open issue in the nematicity 
is the emergence of another type of nematicity in various
Ba122 compounds below $T=T^*$, which is higher than $T_S$
by tens of Kelvin, as shown in Fig. \ref{fig:interference}B.
An actual bulk nematic transition at $T=T^*$ has been reported
in many experimental studies,
such as a magnetic torque study \cite{Kasahara-torque}, an X-ray study \cite{X-ray}, an optical measurement study \cite{Thewalt}, and
a laser
photoemission electron microscope (PEEM) study \cite{Shimojima-PEEM}.
Since the orthorhombicity $(a-b)/(a+b)\ll 0.1$\% below $T^*$ is tiny, an extrinsic origin such as the inhomogeneity of the nematic transition 
temperature $T_S$ due to local uniaxial pressure and randomness was proposed
\cite{Fernandes-122,Fisher-nematic,Dai,Dai2,Wiecki-NMR-broad,Lahiri-surface-smectic}.
On the other hand, $T^*$ seems not to be sensitive to the sample quality
\color{black}and the local strain, \color{black}
and the domain structure of nematicity observed above
$T_S$ is homogeneous \cite{Thewalt,Shimojima-PEEM}. It is noteworthy
that the \color{black}bulk  \color{black} orbital polarization starts to emerge at $T=T^* (>T_S)$ in Ba122 compounds,
according to the recent PEEM study \cite{Shimojima-PEEM}.
In this
paper, we will explain the multistage smectic/nematic transitions: the
smectic order $(\q\ne\bm{0})$ 
at $T=T^*$ and the nematic order $(\q=\bm{0})$ at $T_S$. \color{black} In
this scenario, $T^*$ is given by the intrinsic smectic order free from
the randomness. \color{black}

In contrast to the $B_{1g}$ nematicity in typical Fe-based
superconductors, 
the emergence of $45^\circ$ rotated $B_{2g}$ nematicity in heavily
hole-doped AFe$_2$As$_2$ (A=Cs, Rb) has been reported in Refs.
\cite{Cs122-B2g,RbFe2As2-nematic,Moroni-B2g,Shibauchi-Rb122}, while
Refs. \cite{Bohmer-B2g2,Bohmer-B2g} have reported the absence of the nematic order. As shown in
Fig. \ref{fig:interference}C, the dominant $B_{1g}$ nematicity changes
to the $B_{2g}$ nematicity with doping $x$ in
Ba$_{1-x}$Rb$_{x}$Fe$_2$As$_2$.
As for the mechanism of the $B_{2g}$ nematicity, vestigial nematic order by
using the double-stripe magnetic
configuration was suggested \cite{Valenti-Cs122}. However, no SDW
transition has been observed
\cite{Shibauchi-Rb122,CsFe2As2-HF} in AFe$_2$As$_2$, and the spin
fluctuations are weak around $T_S$ in RbFe$_2$As$_2$ \cite{Rb122-NMR}.
In this paper, we reveal the emergence of $B_{2g}$-symmetry bond order
in AFe$_2$As$_2$.

\color{black}As described above, the variety of nematicity in Fe-based
superconductors is very 
rich.
In order to understand the mechanism of nematic/smectic state and superconductivity,
it is important to explain these nematic/smectic states in the same
theoretical framework. \color{black}

\begin{figure}[!htb]
\includegraphics[width=.7\linewidth]{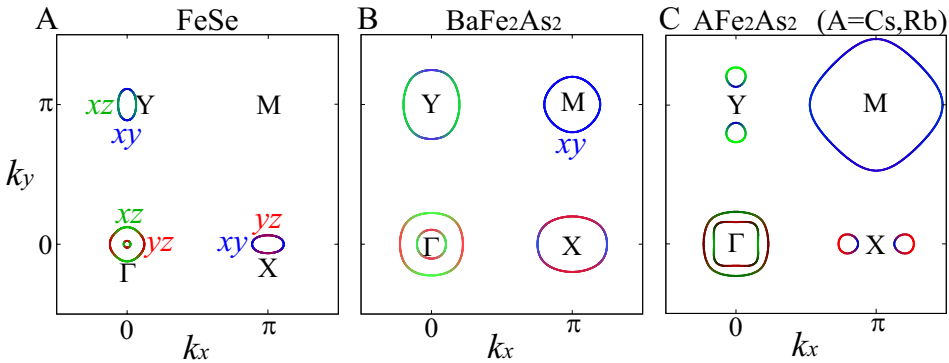}
\caption{
FSs of (A) FeSe $(n_d=6.0)$, (B) BaFe$_2$As$_2$ $(n_d=6.0)$, and (C)
 AFe$_2$As$_2$ (A=Cs, Rb) $(n_d=5.5)$. The colors green, red and blue correspond to orbitals 2, 3, and 4, respectively.
A variety of nematic/smectic states originates from the characteristic structure of FSs.
}
\label{fig:FSs}
\end{figure}

In this paper, we study the $B_{1g}$ nematicity in
FeSe$_{1-x}$Te$_x$ $(n_d=6.0)$, the tiny nematicity below $T^*$ in
BaFe$_2$As$_2$ $(n_d=6.0)$, and the $B_{2g}$ nematicity in AFe$_2$As$_2$
(A=Cs, Rb) $(n_d=5.5)$
by developing the density wave (DW) equation theory. In this theory, the
paramagnon interference mechanism due to the Aslamazov--Larkin (AL) type
VCs shown in
Fig. \ref{fig:interference}D is taken into account.  We also take account of the self-energy effect shown in
Fig. \ref{fig:interference}E.
In this mechanism, the rich variety of nematicity is
naturally understood. \color{black}The obtained nematicity depends on the shape and
topology of FSs, as shown in Figs. \ref{fig:FSs}A, \ref{fig:FSs}B,
\ref{fig:FSs}C. \color{black}
(i) In FeSe$_{1-x}$Te$_x$, all FSs are very small, and $d_{xy}$-orbital hole
pocket is absent. The small spin fluctuations on the three
$d_{xz}$, $d_{yz}$, and $d_{xy}$ orbitals cooperatively lead to the
$B_{1g}$ nematic order, where the orbital order for $d_{xz}$ and
$d_{yz}$ orbitals coexists with the bond order for the $d_{xy}$
orbital. The experimental Lifshitz transition below $T_S$ is naturally
explained by the nematic order.
(ii) In BaFe$_2$As$_2$, the $d_{xy}$ hole pocket emerges, and
each FS is relatively large.
The smectic order at $T=T^*(>T_S)$ and the nematic order at
$T=T_S$ emerge due to the strong $d_{xy}$-orbital nesting.
\color{black}The smectic order explains the tiny nematicity below $T^*$, and the multistage transitions
are explained by the smectic and nematic orders. \color{black}
(iii) In heavily hole-doped AFe$_2$As$_2$, the large $d_{xy}$-orbital hole pocket and
the four tiny Dirac pockets appear. The $B_{2g}$
 nematic bond order emerges due to the $d_{xy}$-orbital paramagnon
 interference mechanism, where the nesting between
the Dirac pockets and the large $d_{xy}$-orbital hole pocket plays an
important role.
\color{black}By considering the fermiology of each compound, these various
nematic/smectic states are explained by the same theoretical framework
base on the
paramagnon interference mechanism. \color{black}

\color{black}In the present study, we intensively study the effect of the
self-energy on the nematic/smectic orders. It has been dropped in
many previous studies, in spite of the fact that the self-energy is
necessary to satisfy the criteria of Baym-Kadanoff's conserving
approximation \cite{Baym,Tremblay}. We revealed that (a) the nematic/smectic order is
stabilized by the AL-type VCs, while (b) $T_S$ is reduced to become realistic
($\sim 100$K) by introducing the self-energy. These results validate the idea
of the "nematic/smectic state due to the paramagnon interference
mechanism" proposed in our previous studies \cite{Onari-SCVC,Onari-Hdope,Onari-form,Yamakawa-PRX,Yamakawa-FeSe-super,Onari-B2g,Onari-122,Kawaguchi,Yamakawa-Cu,Tsuchiizu-Cu,Onari-TBG,Hirata,Tazai-kagome}. In addition,
(c) the phase diagram of FeSe$_{1-x}$Te$_x$
\cite{FeSeTe,FeSeTe2,FeSeTe3} is understood by using a fixed
Coulomb interaction thanks to the self-energy. (In the absence of the
self-energy, add-hoc doping $x$ dependence of the Coulomb interaction has to be introduced.) The
main merits (a)-(c) in the present study strongly indicate that the
nematic/smectic states originate from the paramagnon interference
mechanism \cite{Onari-SCVC,Onari-Hdope,Onari-form,Yamakawa-PRX,Yamakawa-FeSe-super,Onari-B2g,Onari-122,Kawaguchi,Yamakawa-Cu,Tsuchiizu-Cu,Onari-TBG,Hirata,Tazai-kagome}.
\color{black}

\section{Multiorbital models and Formulation}
\subsection{Multiorbital models}
Here, we introduce the multiorbital models based on the first-principles calculations.
\color{black}We analyze the following two-dimensional 
$d$-$p$ Hubbard model with a unique parameter $r$, which controls the
strength of the Coulomb interaction \cite{Onari-form}: \color{black}
\begin{eqnarray}
H_x=H^0_x+rH^U, \ \ \
\label{eqn:Ham}
\end{eqnarray}
where 
$H^0_x$ is the first-principles model, and $H^U$ is the Coulomb interaction for
$d$-orbitals. We neglect the Coulomb interaction for $p$-orbitals.
We denote the five Fe $d$-orbitals $d_{3z^2-r^2}$, $d_{xz}$,
$d_{yz}$, $d_{xy}$, $d_{x^2-y^2}$ as $l=1,2,3,4,5$, and three
Se(As) $p$-orbitals $p_x$, $p_y$, $p_z$ as $l=6,7,8$.
To obtain the model, we first use the WIEN2k \cite{wien2k} and Wannier90
\cite{wannier90} codes.
Next, to reproduce the experimentally observed FSs,
we introduce the $\k$-dependent shifts for orbital $l$, $\delta E_l$,
by modifying the intra-orbital hopping parameters,
as explained in Ref. \cite{Yamakawa-PRX}.
In FeSe$_{1-x}$Te$_x$ model, we shift the $d_{xy}$-orbital band and the $d_{xz/yz}$-orbital band
at [$\Gamma$, M, X] points
by [$0$eV, $-0.27$eV, $+0.40$eV] and [$-0.24$eV, $0$eV, $+0.13$eV],
respectively. In BaFe$_2$As$_2$ model, the shifts are absent.
In CsFe$_2$As$_2$ model, we shift the $d_{xy}$-orbital band and the $d_{xz/yz}$-orbital band
at [$\Gamma$, M, X] points
by [$0$eV, $+0.40$eV, $0$eV] and [$-0.40$eV, $0$eV, $+0.10$eV], respectively.

We employ the $d$-orbital Coulomb interaction introduced by the
 constraint random phase approximation (RPA)
 method in Ref. \cite{Arita}.
The Coulomb interactions for the spin and charge channels are generally given as

\begin{equation}
U^{\mathrm{s}}_{l_{1},l_{2};l_{3},l_{4}} = \begin{cases}
U_{l_1,l_1}, & l_1=l_2=l_3=l_4 \\
U_{l_1,l_2}' , & l_1=l_3 \neq l_2=l_4 \\
J_{l_1,l_3}, & l_1=l_2 \neq l_3=l_4 \\
J_{l_1,l_2}, & l_1=l_4 \neq l_2=l_3 \\
0 , & \mathrm{otherwise}.
\end{cases}
\end{equation}

\begin{equation}
U^{\mathrm{c}}_{l_{1},l_{2};l_{3},l_{4}} = \begin{cases}
-U_{l_1,l_1}, & l_1=l_2=l_3=l_4 \\
U_{l_1,l_2}'-2J_{1_1,l_2} , & l_1=l_3 \neq l_2=l_4 \\
-2U_{l_1,l_3}' + J_{l_1,l_3} , & l_1=l_2 \neq l_3=l_4 \\
-J_{1_1,l_2} , &l_1=l_4 \neq l_2=l_3 \\
0 . & \mathrm{otherwise}.
\end{cases}
\end{equation}

Hamiltonian of the Coulomb interaction is given as
\begin{eqnarray}
H^U&=&-\!\!\!\!\sum_{\k\bm{k'}\q,\sigma\sigma'}\sum_{l_1l_2l_3l_4}\left(\frac{U^c+U^s\sigma\sigma'}{4}\right)_{l_1,l_2;l_3,l_4}\nonumber \\
&&\times c^{l_1\dagger}_{\k+\q,\sigma}c^{l_2}_{\k,\sigma}c^{l_3\dagger}_{\bm{k'}-\q,\sigma'}c^{l_4}_{\bm{k'},\sigma'},
\end{eqnarray}
where $\sigma,\sigma'=\pm 1$ denote spin.

By using the multiorbital Coulomb interaction,
the spin (charge) susceptibility ${\hat \chi}^{s(c)}(q)$ for
$q=(\q,\w_m=2m\pi T)$ is given by  
\begin{equation}
{\hat \chi}^{s(c)}(q)={\hat\chi^0}(q)[1-{\hat U}^{s(c)}{\hat
\chi^0(q)}]^{-1},
\end{equation}
where the irreducible susceptibility is
\begin{equation}
\chi^0_{l,l';m,m'}(q)= -\frac{T}{N}\sum_k
G_{l,m}(k+q)G_{m',l'}(k).
\end{equation}
${\hat G}(k)$ is the multiorbital Green function with the self-energy
$\hat{\Sigma}$ and given as
${\hat G}(k)=[(i\e_n+\mu){\hat1}-{\hat{h}}^0(\k)-\hat{\Sigma}(k)]^{-1}$ 
for $=[\k,\e_n=(2n+1)\pi T]$. 
Here, ${\hat{h}}^0(\k)$ is the matrix expression of $H^0$, 
and $\mu$ is the chemical potential.
The spin (charge) Stoner factor $\alpha_{s(c)}$ 
is defined as
the maximum eigenvalue of $\hat{U}^{s(c)}\hat{\chi}^0(\bm{q},0)$.
Since ${\hat
\chi}^{s(c)}(q)\propto(1-\a_{s(c)})^{-1}$ holds, spin (charge)
fluctuations develop with increasing $\a_{s(c)}$, and $\a_{s(c)}=1$
corresponds to spin- (charge)-channel ordered
state.

\subsection{FLEX approximation}
Here, we introduce the multiorbital
fluctuation exchange (FLEX) approximation \cite{Onari-Hdope,Bickers}.
The FLEX approximation satisfies the
conserving-approximation formalism of Baym and Kadanoff
\cite{Baym,Tremblay}.
In the FLEX approximation, the self-energy is given as
\begin{equation}
\hat{\Sigma}(k)=\frac{T}{N}\sum_{k'}\hat{V}^{\Sigma}(k-k')\hat{G}(k'),
\end{equation}
which is shown by the Feynman diagram in Fig. \ref{fig:interference}E.
The effective interaction $\hat{V}^\Sigma$
for the self-energy in the FLEX approximation is given as
\begin{eqnarray}
\hat{V}^\Sigma(q)&=&
 \frac{3}{2}\hat{U}^{s}\hat{\chi}^{s}(q)\hat{U}^{s}+\frac{1}{2}\hat{U}^{c}\hat{\chi}^{c}(q)\hat{U}^{c}+\frac{3}{2}\hat{U}^s+\frac{1}{2}\hat{U}^c\nonumber
 \\
&&-\hat{U}^{\uparrow\downarrow}\hat{\chi}^0(q)\hat{U}^{\uparrow\downarrow}-\frac{1}{2}\hat{U}^{\uparrow\uparrow}\hat{\chi}^0(q)\hat{U}^{\uparrow\uparrow},
\end{eqnarray}
where $\hat{U}^{\uparrow\downarrow}\equiv \frac{\hat{U}^c-\hat{U}^s}{2}$
and $\hat{U}^{\uparrow\uparrow}\equiv \frac{\hat{U}^c+\hat{U}^s}{2}$
are denoted.
We set $\mu=0$. $\hat{\chi}^{s(c)}(q)$, $\hat{\Sigma}(k)$, and
$\hat{G}(k)$ are calculated self-consistently.
In multiband systems, the FSs are modified from the original FSs
due to the self-energy correction.
To escape from this difficulty, we subtract the Hermite term
$[\hat{\Sigma}(\bm{k},+i0)+\hat{\Sigma}(\bm{k},-i0)]/2$
from the original self-energy, which corresponds to the
elimination of double-counting terms between LDA and FLEX.

\subsection{DW equation}
\begin{figure}[!htb]
\includegraphics[width=.5\linewidth]{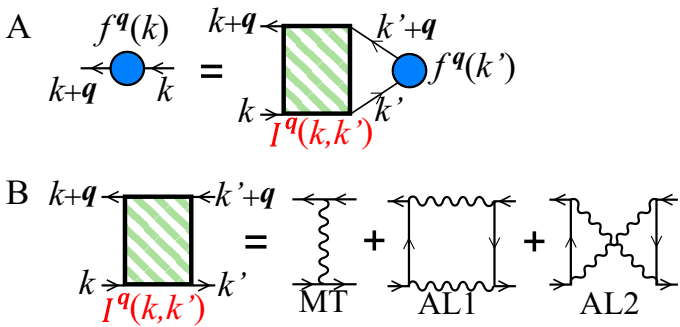}
 \caption{
Feynman diagrams of (A) the DW
 equation and (B) the charge-channel irreducible four point vertex. Each wavy line represents a
 spin-fluctuation-mediated interaction.
}
\label{fig:diagram}
\end{figure}

\color{black}We derive the strongest charge-channel density-wave (DW)
instability without assuming the order parameter and wavevector. \color{black}
For this purpose, we use the DW equation method developed in Refs. \cite{Onari-form,Onari-B2g,Kontani-sLC}. 
We obtain the optimized non-local form factor
$\hat{f}^\q(k)$ with the momentum and orbital dependences by solving the following
 linearized DW equation shown in Fig. \ref{fig:diagram}A:
%
\begin{eqnarray}
  \!\!\!\!\lambda_\q f^\q_{l,l'}(k)&=& \frac{T}{N}
 \!\!\sum_{k',m,m'} {K}^{\bm{q}}_{l,l';m,m'}(k,k')f^\q_{m,m'}(k'),
 \label{eq:DW}\\ 
 \!\!\!\!K^{\bm{q}}_{l,l';m,m'}(k,k')&=& \!\!\!\!\sum_{m_1,m_2} \!\!\!\!I^{\bm{q}}_{l,l';m_1,m_2}(k,k')g^{\bm{q}}_{m_1,m_2;m,m'}(k'),\label{eqn:K}
\end{eqnarray}
where $\lambda_\q$ is the eigenvalue of the form factor $\hat{f}^\q(k)$,
$g^{\bm{q}}_{l,l';m,m'}(k)\equiv
-G_{l,m}\left(k+\bm{q}\right)G_{m',l'}(k)$,
and $\hat{I}^{\bm{q}}(k,k')$ is the charge-channel irreducible
four-point vertex shown in Fig. \ref{fig:diagram}B. 
The four-point vertex interaction $\hat{I}^{\bm{q}}(k,k')$
in the DW equation (\ref{eqn:K}) \cite{Onari-form,Onari-B2g} is given by
\begin{eqnarray}
&& \!\!\!\!\!\!\!\!\!\!\!
I^{\bm{q}}_{l,l';m,m'}(k,k')=\sum_{b=s,c}
\left[-\frac{a^b}{2} V^{b}_{l,m;l',m'}(k-k')\right.
\nonumber \\
&& 
+\frac{T}{N}\!\!\!\!\sum_{p,l_1,l_2,m_1,m_2}\!\!\!\!\!\!\!\!\!\!
 \frac{a^b}{2} V^{b}_{l,l_1;m,m_2}\left(p+{\q}\right)V^{b}_{m',l_2;l',m_1}\left(p\right)
 \nonumber \\
&& \qquad\qquad
\times G_{l_1,m_1}(k-p)G_{l_2,m_2}(k'-p)
\nonumber \\
&&
+\frac{T}{N}\!\!\!\!\sum_{p,l_1,l_2,m_1,m_2}\!\!\!\!\!\!\!\!\!\!
 \frac{a^b}{2} V^{b}_{l,l_1;l_2,m'}\left(p+\q\right)V^{b}_{m_2,m;l',m_1}\left(p\right)
 \nonumber \\
&& \qquad\qquad
\left.\times G_{l_1,m_1}(k-p)G_{l_2,m_2}(k'+p+\q)\right],
\label{eqn:S-K} 
\end{eqnarray}
where $a^s=3$, $a^c=1$, $p=(\p,\w_l)$, and
$\hat{V}^{s(c)}(q)=\hat{U}^{s(c)}+\hat{U}^{s(c)}\hat{\chi}^{s(c)}(q)\hat{U}^{s(c)}$.

In Eq. (\ref{eqn:S-K}),
the first line corresponds to the Maki-Thompson (MT) term,
and the second and third lines give the AL terms,
respectively. Feynman diagrams of the MT terms and AL terms are shown in
Fig. \ref{fig:diagram}B.

The AL terms are enhanced by the paramagnon interference
$\hat{\chi}^{s}(\Q)\times\hat{\chi}^{s}(\bm{Q'})$ shown in Fig
\ref{fig:interference}D. Thus, $\q=\Q+\bm{Q'}=\bm{0}$ nematic order
is naturally induced by the
paramagnon interference at the same nesting vector
$(\bm{Q'}=-\Q)$.
In the MT term,
the first-order term with respect to ${\hat{U}}^{s,c}$ 
gives the Hartree--Fock (HF) term in the mean-field theory.
The charge-channel DW with wavevector $\q$ is established when
the largest $\lambda_\q=1$. Thus, the smaller $\lambda_\q$ corresponds
to the lower $T_S$.
The DW susceptibility is proportional to
 $1/(1-\lambda_\q)$ as explained in Ref.\cite{Onari-122}. Therefore, $\lambda_\q$ represents the strength
 of the DW instability.


\section{Results and discussions}
\subsection{Results of FeSe$_{1-x}$Te$_x$}
In this section, we show that (i) the $B_{1g}$ nematic orbital+bond order
originates from the paramagnon interference, and (ii) the effect of self-energy is essential to
reproduce the $x$ dependence of $T_S$ shown in
Fig. \ref{fig:interference}A in FeSe$_{1-x}$Te$_x$. \color{black}
The effect of the
self-energy on the nematic/smectic order caused by the VCs is
systematically studied in the present manuscript. Thanks to the
self-energy, $T_S$ is reduced to become realistic ($\sim100$K), while the
symmetry of the nematic/smectic order is unchanged. Thus, the idea of
electronic nematicity due to "the paramagnon-interference mechanism"
proposed in Refs. \cite{Onari-SCVC,Onari-Hdope,Onari-form,Yamakawa-PRX,Yamakawa-FeSe-super,Onari-B2g,Onari-122,Kawaguchi,Yamakawa-Cu,Tsuchiizu-Cu,Onari-TBG,Hirata,Tazai-kagome} has been confirmed by the present
study. \color{black}
 \color{black}
\color{black}Hereafter, we fix $r=0.35$, $T=15$meV in calculations with the
self-energy and $r=0.15$, $T=15$meV in calculations without the self-energy, unless otherwise noted. \color{black}

\begin{figure}[!htb]
\includegraphics[width=.8\linewidth]{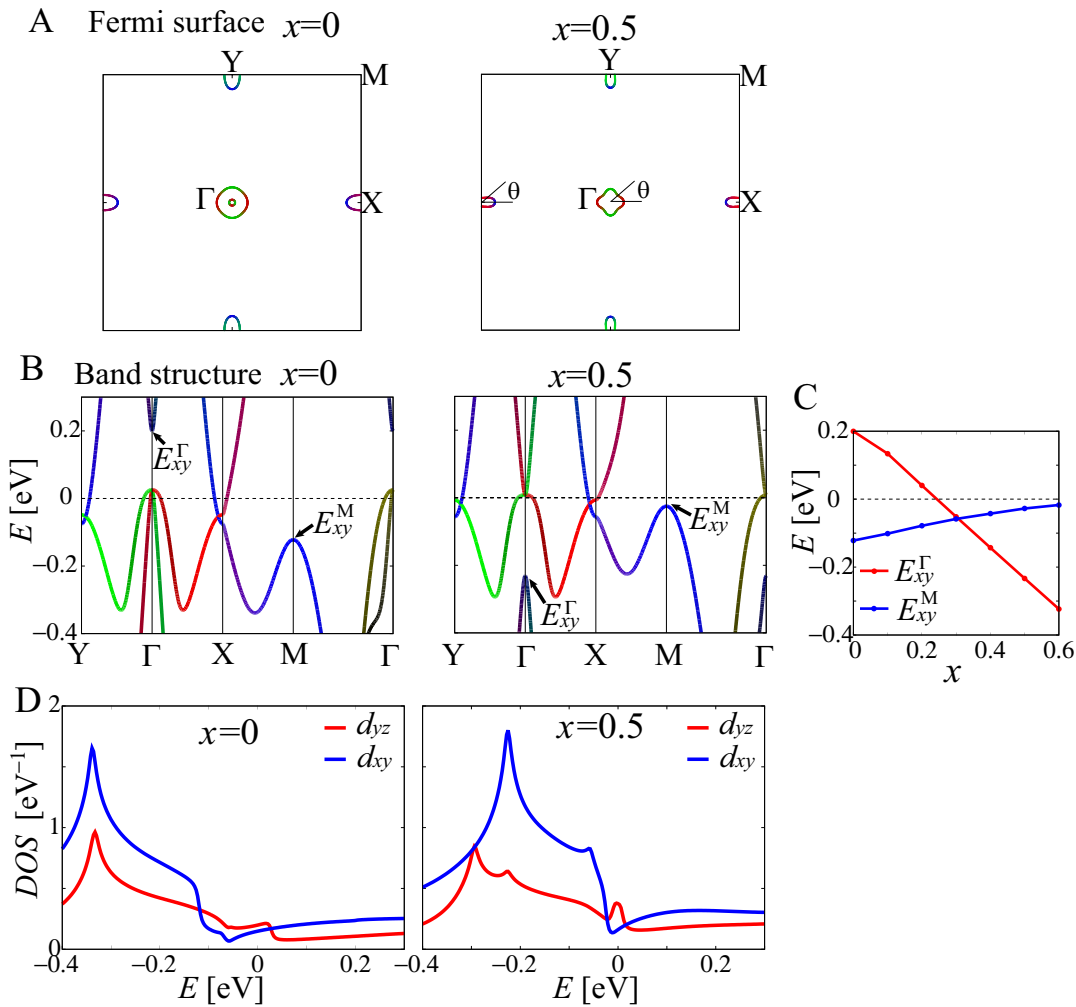}
\caption{
(A) FSs and (B) band structures of FeSe$_{1-x}$Te$_x$ for $x=0$ and $0.5$.
 (C) $x$ dependences of $E_{xy}^\Gamma$ and $E_{xy}^{\rm M}$.
(D) DOS of orbitals 3 and 4 for $x=0$, $0.5$.}
\label{fig:FS}
\end{figure}

Figures \ref{fig:FS}A and \ref{fig:FS}B show $x$ dependent
FSs and band structures, respectively. 
The FSs are small compared to other Fe-based superconductors.
\color{black} $d_{xy}$ orbital level $E_{xy}^M$ at M
 point increases with increasing $x$, as shown in Fig. \ref{fig:FS}C. \color{black}
This behavior is consistent with ARPES measurements
\cite{FeSeTe-ARPES,FeSeTe-ARPES2}. 
On the other hand, $d_{xy}$ orbital level $E_{xy}^\Gamma$ at $\Gamma$
 point decreases with increasing $x$.
 $E_{xy}^\Gamma$ becomes lower than the $d_{xz(yz)}$ orbital level for
 $x\gtrsim 0.3$, and the topology of band changes.
 The change of topology has been observed between $\Gamma$ and $Z$
 points in ARPES measurements of
 FeSe$_{0.5}$Te$_{0.5}$ \cite{FeSeTe-Topo,FeSeTe-Topo2}.
Figure \ref{fig:FS}D shows the density of state (DOS) of orbitals 3 and
4 for $x=0$, $0.5$. DOS near the Fermi level for $x=0.5$ is larger than
that for $x=0$ since the bandwidth decreases, and $E_{xy}^M$ comes close
to the Fermi level with increasing $x$. In addition, the dispersion of
orbitals 2 and 3 at
$\Gamma$ point becomes flat as $E_{xy}^\Gamma$ decreases with increasing
$x$, which also enlarges the DOS for orbitals 2 and 3 near the Fermi level.

\begin{figure}[!htb]
\includegraphics[width=.35\linewidth]{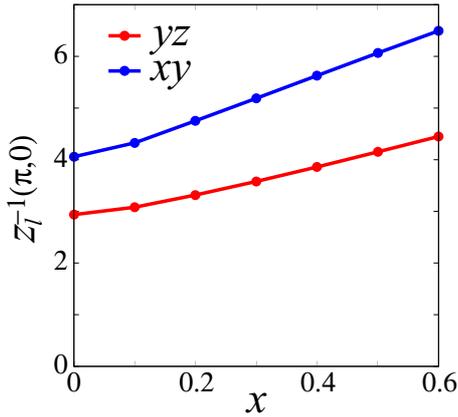}
 \caption{
$x$ dependences of mass enhancement factor $z^{-1}_l(\pi,0)$ for orbitals
 $l=3$ and $4$.
}
\label{fig:Sigma}
\end{figure}

In order to discuss the self-energy effect, we calculate the mass
enhancement factors.
Figure \ref{fig:Sigma} shows the obtained $x$ dependence
of the mass
enhancement factors $z^{-1}_l(\pi,0)$ for orbital $l=3,4$, which are 
given by $z^{-1}_l(\k)=1-{\rm Im}{\Sigma_{l,l}(\k,\pi T)}/{\pi T}$ in
the FLEX approximation. 
\color{black}The value of $z^{-1}_l(\pi,0)$ increases with increasing $x$ since the
electron correlation increases due to the reduction of bandwidth and the increase of DOS shown in Fig. \ref{fig:FS}D. Particularly, $z^{-1}_4(\pi,0)$ is enhanced by the $d_{xy}$ orbital electron correlation between the electron pockets and the band around M point since $E_{xy}^{\rm M}$ comes close to the Fermi
level, as shown in Fig. \ref{fig:FS}C. \color{black}
The behaviors of $z^{-1}_l$ are similar to those given by the
dynamical mean-field theory \cite{DMFT} and experiment \cite{mass1}.

\begin{figure}[!htb]
\includegraphics[width=.4\linewidth]{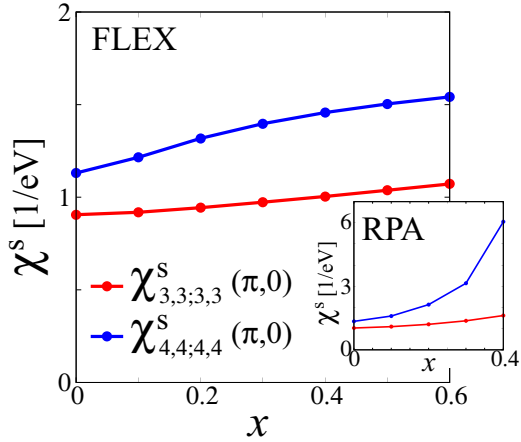}
 \caption{
 $x$ dependences of $\chi^s_{3,3;3,3(4,4;4,4)}(\pi,0)$ in the FLEX
 approximation. Those in the RPA are shown in the inset.
}
\label{fig:chis}
\end{figure}

Figure \ref{fig:chis} shows $x$ dependences of
$\chi^s_{3,3;3,3}(\pi,0)$ and $\chi^s_{4,4;4,4}(\pi,0)$ in the FLEX
approximation and the RPA.
$\chi^s_{3,3;3,3}(\pi,0)$ is almost independent of doping $x$, which
means that change of topology or number of FS around $\Gamma$ composed of $d_{xz}$ and $d_{yz}$
orbitals does not strongly affect the spin fluctuation for $d_{xz(yz)}$
orbital.
On the other hand, $\chi^s_{4,4;4,4}(\pi,0)$ in the RPA without the
self-energy is strongly enhanced with increasing
$x$ since the electron correlation for $d_{xy}$
orbital between electron pockets and the band around M point is significant
for the enhancement of $\chi^s_{4,4;4,4}(\pi,0)$.
The strong enhancement of $\chi^s_{4,4;4,4}(\pi,0)$ is suppressed by the
self-energy in the FLEX approximation. 
This suppression is necessary to reproduce the $x$ dependence of $T_S$ in the phase
diagram.

\begin{figure}[!htb]
\includegraphics[width=.9\linewidth]{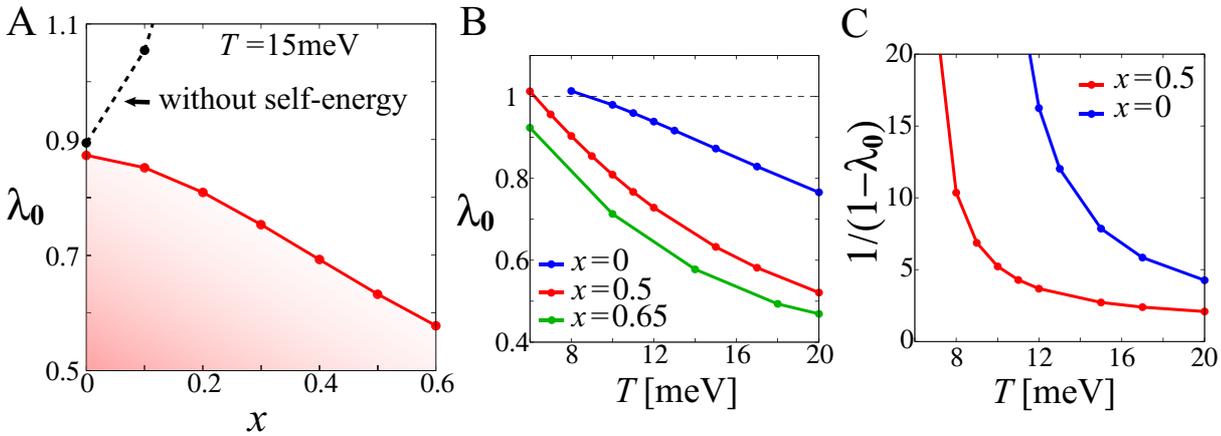}
 \caption{
(A) $x$ dependences of $\lambda_{\bm{0}}$
 in the DW equation with and without the self-energy at $T=15$meV. 
(B) $T$ dependences of $\lambda_{\bm{0}}$ with the self-energy for
 $x=0,0.5,0.65$.
(C) $T$ dependences of $1/(1-\lambda_{\bm{0}})$ with the self-energy for $x=0,0.5$.
}
\label{fig:xdep}
\end{figure}

Hereafter, we discuss the DW instability given by the DW equation (\ref{eq:DW}).
Figure \ref{fig:xdep}A shows $x$ dependences of $\lambda_{\bm{0}}$ for
the $B_{1g}$ nematic state with and without the self-energy.
$\lambda_{\bm{0}}$ without the self-energy rapidly increases with doping $x$ due to 
the paramagnon interference shown in Fig. \ref{fig:interference}D. 
$\lambda_{\bm{0}}$ is enlarged by the interference
between $\chi^s_{4,4;4,4}$ strongly enhanced in the RPA, as shown in Fig. \ref{fig:chis}. Since this result
means $T_S$ increases with $x$, the phase diagram in Fig. \ref{fig:interference}A cannot be explained when the self-energy is absent.
However, $\lambda_{\bm{0}}$ including the self-energy decreases with
doping $x$ since the enhancement of $\chi^s_{4,4;4,4}$ in the FLEX
approximation is moderate and
the self-energy suppresses the $\hat{G}$ and $\hat{I}$ in the DW equation
 (\ref{eqn:K}). \color{black}The value of $\lambda_{\bm{0}}$ increases with decreasing $T$,
as shown in Fig. \ref{fig:xdep}B, $T=T_S$ is given when
$\lambda_{\bm{0}}=1$ is satisfied. \color{black} Thus, $T_S$ at $x=0$ is higher than
that at $x=0.5$, and $T_S$ at $x=0.65$ cannot be obtained for $T>6$meV.
The $x$ dependence of $T_S$ obtained by the paramagnon
interference mechanism is consistent
with the phase diagram in Fig. \ref{fig:interference}A \cite{FeSeTe}.
We see that $T$ dependences of the strength of nematic fluctuations
$1/(1-\lambda_{\bm{0}})$ satisfy the Curie--Weiss law at low
temperatures, as shown in Fig. \ref{fig:xdep}C.
We note that the $B_{1g}$ nematic state is realized due to the small FSs
even for the weak spin fluctuations \cite{Onari-form,Yamakawa-PRX}.

\color{black}Here, we analytically explain that $T_S$ is reduced by the
self-energy by focusing on the
mass renormalization factor $z$. As discussed in Ref. \cite{Yamakawa-PRX}, $\alpha_{s(c)}$ is
independent of $z$ under the scaling $T\rightarrow zT$ and $r\rightarrow
r/z$. Under this scaling, the eigenvalue of the DW equation is unchanged \cite{Yamakawa-PRX}. Thus, $T_S$ obtained by the DW equation without the self-energy is
reduced to $zT_S$ due to the self-energy. As a result, realistic $T_S$ is
obtained by taking the self-energy into account.
\color{black}

\begin{figure}[!htb]
\includegraphics[width=.6\linewidth]{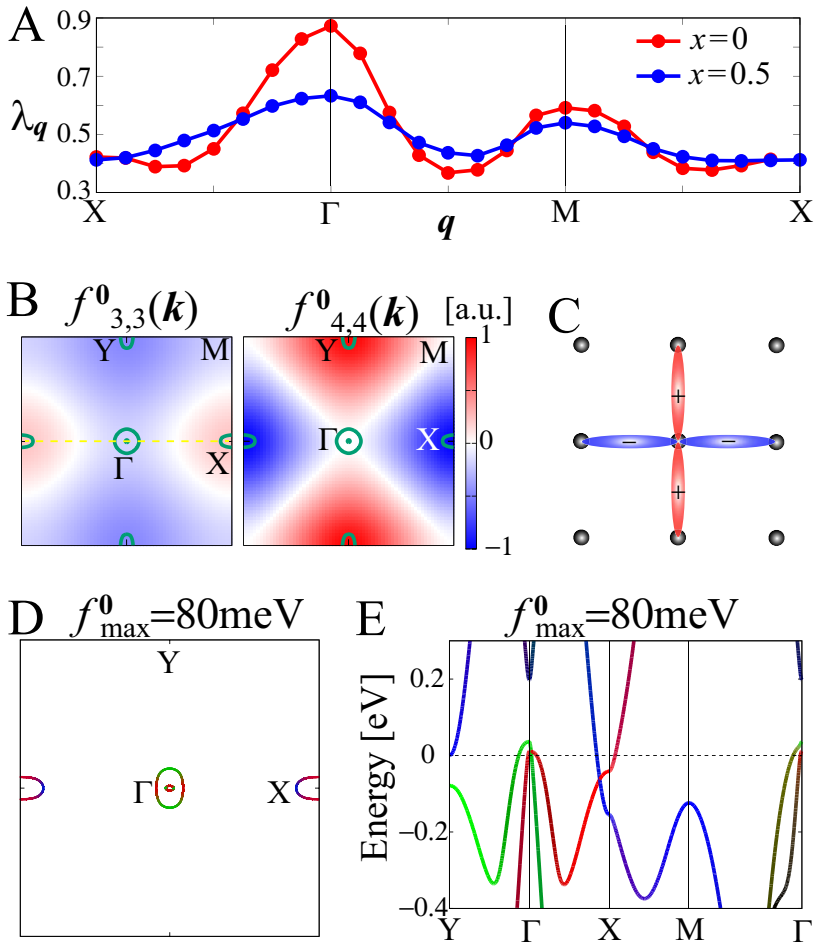}
\caption{
(A) $\q$ dependences of $\lambda_{\q}$ with the self-energy for $x=0$, $0.5$.
(B) $\k$ dependences of $f^{\bm{0}}_{3,3}(\k)$ and
 $f^{\bm{0}}_{4,4}(\k)$ for $x=0$, where green lines denote FSs.
$f^{\bm{0}}_{3,3}(\k)$ changes sign along $k_x$ axis (yellow dashed
 line).
(C) $B_{1g}$ nearest-neighbor bond order corresponding to $f^{\bm{0}}_{4,4}(\k)$.
(D) FSs and (E) band structure under the nematic order with $f^{\bm{0}}_{\rm max}=80$meV for $x=0$.
}
\label{fig:form}
\end{figure}

\color{black}Figure \ref{fig:form}A shows $\q$ dependences of
$\lambda_{\q}$ with the self-energy at $x=0$, $0.5$. \color{black} $\lambda_{\q}$ has peak at $\q=\bm{0}$, which
means that the ferro nematic order is favored.
Figure \ref{fig:form}B shows $\k$ dependences of the static
form factors $f^{\bm{0}}_{33}(\k)$ and $f^{\bm{0}}_{44}(\k)$, where $\hat{f}^{\q}(\k)$ is given by the
analytic continuation of $\hat{f}^\q(k)$.
$f^{\bm{0}}_{33}(k_x,k_y)=-f^{\bm{0}}_{22}(k_y,k_x)$ represents $B_{1g}$
orbital
order between orbitals 2 and 3. From the $\k$ dependence of
$f^{\bm{0}}_{33(22)}(\k)$, the sign-reversing orbital
order is confirmed along $k_x(k_y)$ axis. As shown in Fig. \ref{fig:form}C, $\k$ dependence of
$f^{\bm{0}}_{44}(\k)\propto \cos(k_x)-\cos(k_y)$ causes the B$_{1g}$
nearest-neighbor bond
order, which is the modulation of correlated hopping.
Based on the paramagnon interference
mechanism, we find that the small spin fluctuations on the three $d_{xz}$,
$d_{yz}$, and $d_{xy}$ orbitals cooperatively cause the $B_{1g}$ nematic
orbital+bond order.
\color{black}The FSs and the band structure under the nematic order with the maximum
value of the form factor $f^{\bm{0}}_{\rm max}=80$meV are
shown in Figs. \ref{fig:form}D and \ref{fig:form}E. $f^{\bm{0}}_{\rm
max}=80$meV with the mass enhancement factor
$z^{-1}_l=2\sim 4$ is consistent with ARPES measurements
\cite{FeSe-ARPES-Shen,FeSe-ARPES-Shen2}. \color{black}
The Lifshitz transition, where the FS around Y point is missing, has been reported in recent experiments
\cite{FeSe-Y,FeSe-Y2,FeSe-Y4,FeSe-Y3}. The Lifshitz transition is
naturally explained by
the increase of $d_{xy}$ level around Y point induced by $f^{\bm{0}}_{44}(\k)$.
\color{black}We note that the obtained coexistence of the bond order on
the $d_{xy}$ orbital and the orbital order on the ($d_{xz}$, $d_{yz}$)
orbitals has already been shown in the supplementary material of
Refs. \cite{Onari-B2g,Onari-122}. \color{black}
In Fig. \ref{fig:form}D, we derived the Lifshitz transition by setting
$f^{\bm{0}}_{\rm max}=80$meV by hand. It is noteworthy that the same
result is recently obtained by solving the full DW equation in
Ref. \cite{Tazai-LW}. \color{black}The full DW equation enables us to study the
electronic states below $T_S$ without introducing additional fitting
parameters. \color{black}

\begin{figure}[!htb]
\includegraphics[width=.6\linewidth]{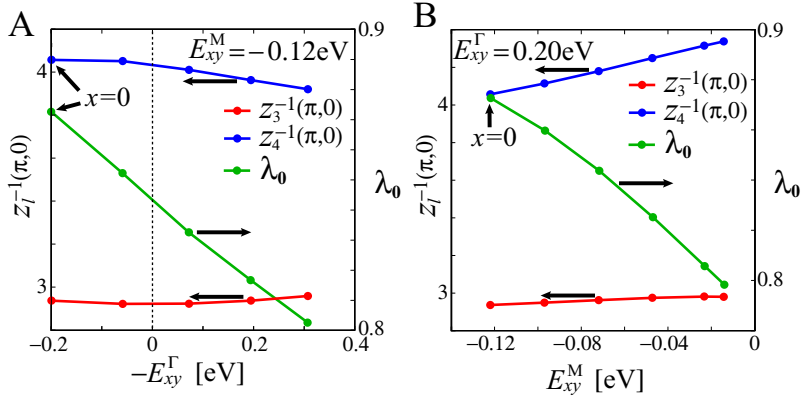}
\caption{
(A) $E_{xy}^{\Gamma}$ dependences of  $z^{-1}_l(\pi,0)$ for $l=3,4$, and
 $\lambda_{\bm{0}}$ given by introducing only $E_{xy}^{\Gamma}$ shift for
 the $x=0$ model.
(B) $E_{xy}^{\rm M}$ dependences of $z^{-1}_l(\pi,0)$ for $l=3,4$, and
 $\lambda_{\bm{0}}$ given by introducing only $E_{xy}^{\rm M}$ shift for
 the $x=0$ model. 
}
\label{fig:DeltaExy}
\end{figure}

Here, we confirm that the $d_{xy}$ orbital levels at $\Gamma$ and M
points are important for the $x$ dependence of
$\lambda_{\bm{0}}$.
We employ the simple model, where only the shift of $E_{xy}^{\Gamma}$ or
$E_{xy}^{\rm M}$ is introduced for the $x=0$ model.
Figure \ref{fig:DeltaExy}A shows $E_{xy}^{\Gamma}$ dependences of $z^{-1}_l(\pi,0)$ and
$\lambda_{\bm{0}}$, respectively.
 $z^{-1}_l(\pi,0)$ is almost independent of the value of $E_{xy}^{\Gamma}$.
$\lambda_{\bm{0}}$ decreases with decreasing $E_{xy}^{\Gamma}$, which is
consistent with the result in Fig. \ref{fig:xdep}A. \color{black}The topology of band structure changes
at $\Gamma$ point with decreasing
$E_{xy}^{\Gamma}$, which plays important role to decrease
$\lambda_{\bm{0}}$. \color{black}
Figure \ref{fig:DeltaExy}B shows $E_{xy}^{\rm M}$ dependences of $z^{-1}_l(\pi,0)$ and
$\lambda_{\bm{0}}$, respectively.
The behaviors of $z^{-1}_4(\pi,0)$ and $\lambda_{\bm{0}}$ are similar to the results in Figs. \ref{fig:Sigma} and \ref{fig:xdep}A.
The $x$ dependences of $z^{-1}_4(\pi,0)$ and $\lambda_{\bm{0}}$ are explained by the
electron correlation between the electron pockets and the $d_{xy}$ band around M
point.
$\lambda_{\bm{0}}$ is suppressed by the self-energy for the
$d_{xy}$ orbital. The suppression becomes strong with increasing $E_{xy}^{\rm M}$ due to the feedback effect of the self-energy.
To summarize, the $B_{1g}$ nematic orbital+bond order is explained by the
paramagnon interference mechanism in FeSe$_{1-x}$Te$_x$, and $x$ dependence of $T_S$ is
well reproduced by the self-energy effect for the $d_{xy}$ orbital.

\subsection{Results of BaFe$_2$As$_2$}
In this section, we discuss the multi nematicity in
BaFe$_2$As$_2$ \cite{Onari-122}. 
\color{black}
The effect of the self-energy on the nematic/smectic orders caused by the VCs is studied in the present manuscript. Transition temperatures are reduced to become realistic due to the self-energy, while the symmetries of the nematic/smectic orders are unchanged. \color{black}
We reveal the origin of tiny nematicity below
$T=T^*$ and explain the multistage transitions 
at $T=T^*$ and $T_S$ in the phase diagram shown in Fig. \ref{fig:interference}B.
\color{black}As shown in Fig. \ref{fig:FSs}B, the size of hole FS around M point
composed of $d_{xy}$ orbital is
similar to that of electron FSs around X and Y points, which causes the good
intra- and inter-orbital nestings. \color{black}
As explained later, the inter-orbital nesting is important
to realize the smectic state at $T=T^*$.

\begin{figure}[!htb]
\includegraphics[width=.6\linewidth]{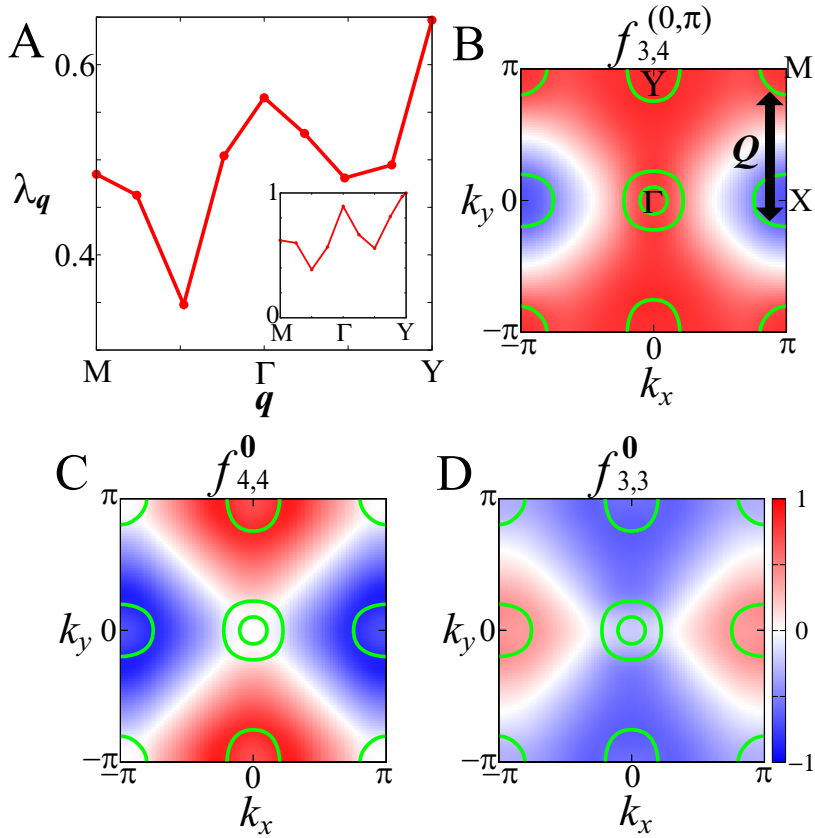}
 \caption{
 (A) $\q$ dependence of $\lambda_{\q}$ with the self-energy for $r=0.68$
 at $T=5$meV
 in BaFe$_2$As$_2$, and that
 without the self-energy for $r=0.30$ at $T=32.4$meV in the inset.
(B) $\k$ dependence of the dominant form factor at $\q=(0,\pi)$,
 $f_{3,4}^{\q}(\k)$ with the
 self-energy,
 which is given by the off-diagonal orbitals 3 and 4.
 $\k$ dependences of form factors at $\q=\bm{0}$, (C)
 $f_{4,4}^{\bm{0}}(\k)$ and (D) $f_{3,3}^{\bm{0}}(\k)$ with the
 self-energy. Green lines denote FSs.
}
\label{fig:DW-Ba122}
\end{figure}

Figure \ref{fig:DW-Ba122}A shows the $\q$-dependence of $\lambda_\q$ with
 and without the self-energy.
$\q=(0,\pi)$ smectic bond order is dominant over the $\q=\bm{0}$ nematic
orbital+bond order because of the relation
$\lambda_{(0,\pi)}>\lambda_{\bm{0}}$, which is robust in the presence of
moderate spin fluctuations $\a_s\gtrsim 0.85$.
Thus, the nematic orbital+bond transition temperature $T_S$ is lower than
$T^*$, where the smectic bond order appears.
Figure \ref{fig:DW-Ba122}B shows
the dominant component of the static form factor, $f^{\q}_{3,4}(\k)$, for
$\q=(0,\pi)$.
Focusing on the X and M points, $f^{(0,\pi)}_{3,4}(\k)$ is proportional to
$-\cos(k_y)$, which corresponds to the inter-orbital smectic bond order,
where the $y$-direction hoppings between orbitals 3 and 4 are
modulated by the correlated hopping $\delta t_{3,4}(y;y\pm1)=-\delta
t_{4,3}(y;y\pm1)=\delta t(-1)^y$. Note that $\delta t_{l,m}(y;y')$ is
real and equal to $\delta
t_{m,l}(y';y)$.

As shown in Fig. \ref{fig:interference}D, the origin of the smectic bond order $f_{3,4}^{(0,\pi)}$ is the quantum
interference between the spin fluctuations
$\chi^s(\Q)$ for $\Q\approx (0,\pi)$ and
$\chi^s(\bm{0})$ due to the AL terms. In this case, $\q=(0,\pi)(=\Q+\Q')$ is
given by $\Q'=\bm{0}$.
$\chi^s(\Q)$ is enhanced when the FS appears around M point since the nesting
between FSs around X and M points becomes good, while the moderate $\chi^s(\bm{0})$ is
caused by the forward scattering.
We find that $f_{3,4}^{(0,\pi)}$ is significantly enlarged by
inter-orbital nesting between the $d_{xy}$-orbital FS around M point and
$d_{yz}$-orbital FS around X point.
In addition to the quantum interference due to the AL terms, the MT terms 
strengthen the sign change of
$f^{(0,\pi)}_{3,4}(\k)$ between X and M points, as reported previously \cite{Onari-form,Onari-B2g,Chubukov-RG1}. Thus, the 
smectic bond order originates from the cooperation between the AL and
MT terms due to the good inter-orbital nesting between FSs around X and M points.
In contrast, the $B_{1g}$ nematic orbital+bond order shown in
Figs. \ref{fig:DW-Ba122}C and \ref{fig:DW-Ba122}D
originates from the interference between $\chi^s(\Q)$
and $\chi^s(-\Q)$.
This nematic orbital+bond order is similar to that in FeSe and
FeSe$_{1-x}$Te$_x$.

\begin{figure}[!htb]
\includegraphics[width=.7\linewidth]{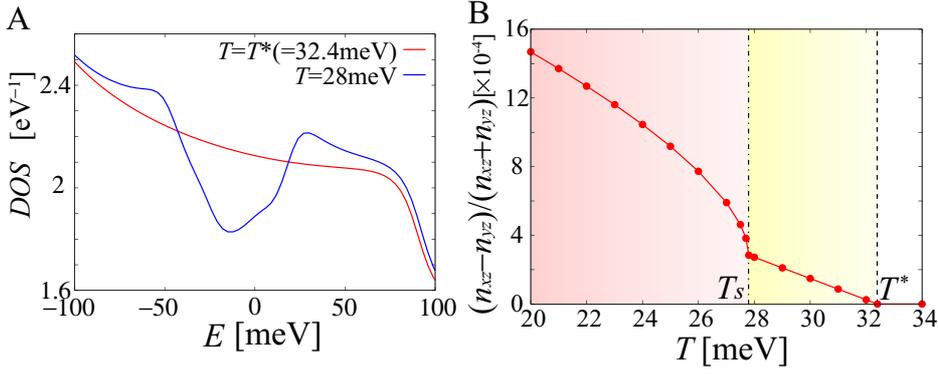}
\caption{
(A) DOS at $T=T^*=32.4$meV and that at $T=28$meV$(<T^*)$ including the
 smectic bond order.
(B) $T$ dependence of nematicity $\psi=(n_2-n_3)/(n_2+n_3)$ including both
  smectic-bond order for $T<T^*$ and ferro-orbital/bond order for
 $T<T_S$.
}
\label{fig:nematicity-Ba122}
\end{figure}

Here, we examine the DOS under the smectic bond order to verify the
present theory. For $T<T^*=32.4$meV without the self-energy, we introduce the
mean-field-like $T$-dependent form factor
$\hat{f}^\q(T)=f^{\rm
max}\tanh\left(1.74\sqrt{T^*/T-1}\right)\hat{f}^{\q}$, where
$\hat{f}^\q$ is the obtained form factor for $\q=(0,\pi)$ normalized as
$\max_{\k} |f^\q(\k)|=1$.
We put $f^{\rm max}=60$meV.
Figure \ref{fig:nematicity-Ba122}A shows the DOS at $T=T^*$ and $28$meV$(<T^*)$.
For $T<T^*$, a pseudogap appears due to the smectic bond order, which is
consistent with the experiments \cite{Moon-PG,Shimojima-PG}. Since the smectic bond order is
an antiferroic order, the folded band structure emerges below $T^*$,
which is also consistent with the experiment \cite{Antiferroic-Ba122}.

Next, we focus on another mystery, the $T$-linear behavior of tiny nematicity $\psi$ in Ba122
\cite{Kasahara-torque} below $T^*$. In order to solve this mystery, we calculate
 the $T$ dependence of uniform nematicity $\psi=(n_2-n_3)/(n_2+n_3)$ in Fig. \ref{fig:nematicity-Ba122}B, where
both $\hat{f}^{(0,\pi)}(T)$ for $T<T^*$ and
the ferro nematic orbital+bond order $\hat{f}^{\bm{0}}(T)$ 
for $T<T_S(=27.8$meV) are introduced.
For $T<T_S$, we assume
$\hat{f}^{\bm{0}}(T)=f^{\rm
max}\tanh\left(1.74\sqrt{T_S/T-1}\right)\hat{f}^{\bm{0}}$,
where $\hat{f}^{\bm{0}}$ is the 
obtained form factor normalized as $\max_{\k}|f^{\bm{0}}(\k)|=1$.
We employ $f^{\rm max}=60$meV, which corresponds to the $d_{xz(yz)}$
orbital energy split $\sim 60$meV in
the ARPES measurements \cite{ARPES} by considering the mass enhancement
factor $z^{-1}_l\sim 2$ for $l=2,3$.
The $T$-linear behavior $\psi\propto(T^*-T)$ for $T_S<T<T^*$ is a consequence of the relation
$\psi\propto[f^{(0,\pi)}(T)]^2$ because the $f^{(0,\pi)}$ term cannot
contribute to any $\q=\bm{0}$ linear response.
Note that the form factor $\hat{f}^{(\pi,0)}$ for
$\q=(\pi,0)$ gives $\psi<0$.
Thus, the $T$-linear behavior of $\psi$ below $T^*$ is also naturally
explained by the smectic bond order.
On the other hand, $\psi\propto\sqrt{T_S-T}$ for $T<T_S$ is
induced by the nematic orbital+bond order. To summarize, the multistage transitions 
at $T=T^*$ and $T_S$, and the $T$-linear $\psi$ below
$T^*$, are naturally explained by the smectic bond order and
nematic orbital+bond order. The hole pocket
around M point is necessary to realize the smectic bond order by the
paramagnon interference mechanism. 

\color{black}We stress that the present
mechanism of the bulk nematicity for $T_S<T<T^*$ is intrinsic and free
from the strength of the disorder and local strain in the
system.  The present smectic order originates from the AL-VC and the FS nesting between the $d_{xy}$-orbital hole pocket and the electron pockets \cite{Onari-122}. We stress that the present theory explains the absence of the smectic order in bulk FeSe \cite{Shimojima-PEEM} because the $d_{xy}$-orbital hole-pocket, which is necessary for the smectic order formation, is below the Fermi level in FeSe.

Here, we explain the details of the recent microscopy measurements in
P-doped Ba122 \cite{Thewalt,Shimojima-PEEM} that support the present intrinsic
scenario. These are bulk and real-space measurements. In the PEEM measurement \cite{Shimojima-PEEM}, very uniform bulk nematic domains have been
observed for $T_S<T<T^*$. The width of each nematic domain
is about $500$nm. The structure of the nematic domains is
unchanged for $T<T_S$. In addition, once the nematic domain completely
disappears by increasing $T$, it never appears at the same location if the
temperature is lowered again. These results are consistent with the
present intrinsic smectic order scenario for $T_S<T<T^*$ in P-doped Ba122. In the photomodulation measurement \cite{Thewalt}, uniform nematic
domains have also been observed. The observed
nematicity becomes small near the nematic domain boundary, irrespective of the fact that the large local strain anisotropy is observed at the domain boundary. The observed anticorrelation between the
nematicity and the local strain anisotropy may conflict with the assumption of the extrinsic scenario of the nematicity above $T_S$.

In contrast, the extrinsic scenario has been proposed by other groups \cite{Fernandes-122,Fisher-nematic,Dai,Dai2,Wiecki-NMR-broad,Lahiri-surface-smectic}. In the extrinsic mechanism, the nematicity for $T>T_S$ in Co-doped Ba122, which exhibits large residual resistivity $(>100\mu\Omega\rm{cm})$, has been explained by the inhomogeneity of $T_S$ induced by the disorder and local strain. However, it is not easy to explain the nematicity above $T_S$ in clean P-doped (non-doped) Ba122 on the same footing in the extrinsic scenario.
\color{black}

 We note that the multistage smectic/nematic
transitions observed in NaFeAs \cite{NaFeAs-T*} are also explained by
the present intrinsic mechanism. \cite{Onari-122}.

\subsection{Results of Ba$_{1-x}$Cs$_x$Fe$_2$As$_2$}
In this section, we discuss the $B_{2g}$ nematicity in
 heavily hole-doped compound AFe$_2$As$_2$ (A=Cs, Rb)
 \cite{Onari-B2g}. 
\color{black}The effect of the
self-energy on the nematic order caused by the VCs is studied in the present manuscript. Thanks to the
self-energy, $T_S$ is reduced to become realistic, while the
symmetry of the nematic order is unchanged. \color{black}
The direction of
$B_{2g}$ nematicity is rotated by $45^\circ$ from that of the conventional $B_{1g}$ nematicity.
Figure \ref{fig:FSs}C shows FSs of CsFe$_2$As$_2$: The hole FS around M
 point composed of $d_{xy}$-orbital is large, while the Dirac pockets near X
and Y points are small. In this system, the $d_{xy}$-orbital spin
 fluctuations are dominant.

\begin{figure}[!htb]
\includegraphics[width=.5\linewidth]{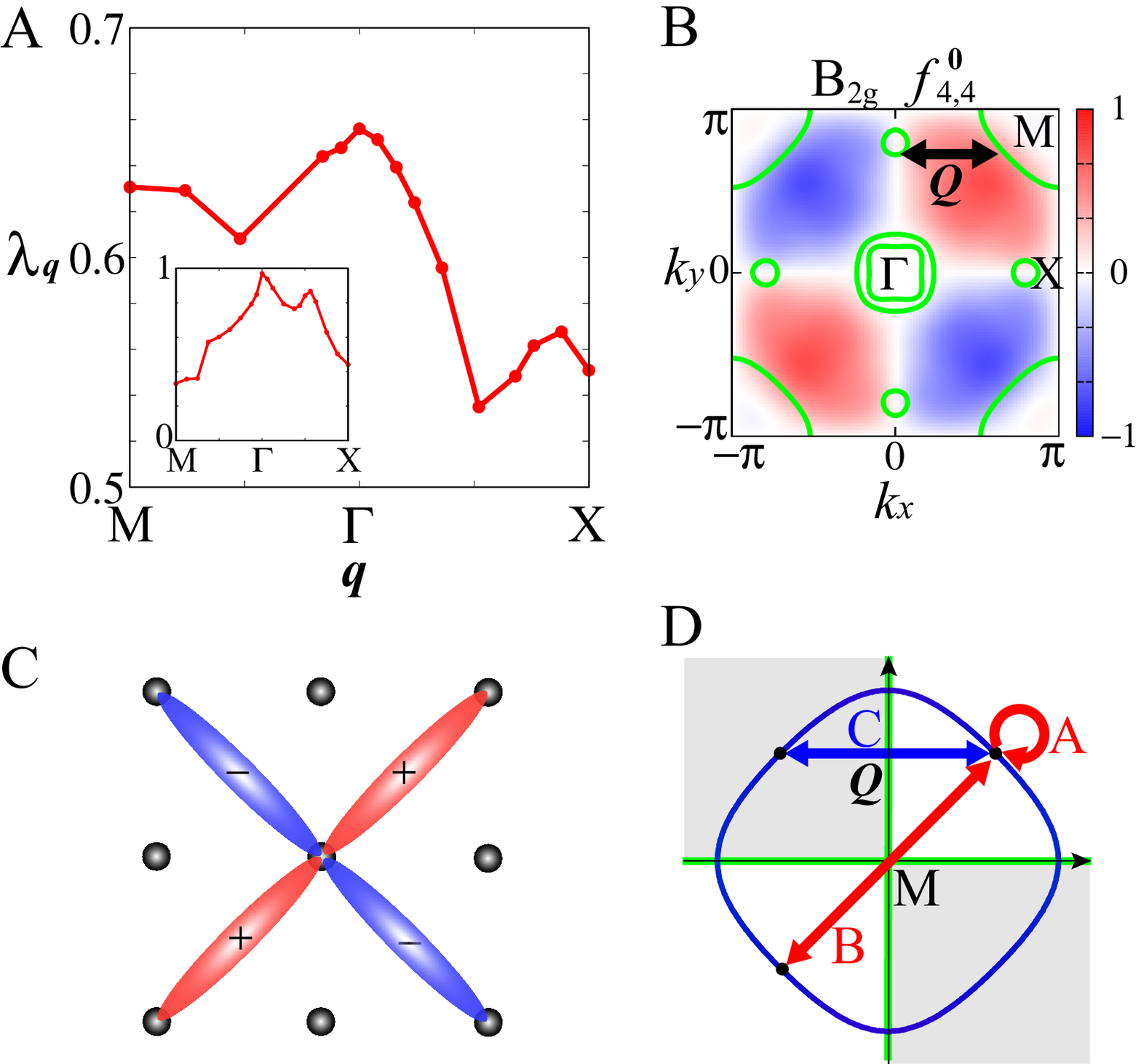}
\caption{
(A) $\q$ dependence of $\lambda_{\q}$ with the self-energy in CsFe$_2$As$_2$, and that without the self-energy in the inset.
(B) $\k$ dependence of $B_{2g}$ form factor
 $f^{\bm{0}}_{4,4}(\k)\propto\sin(k_x)\sin(k_y)$, where the green lines and black
 arrow denote FSs and nesting vector $\Q\sim(0.5\pi,0)$, respectively.
(C) The $B_{2g}$ next-nearest-neighbor bond order corresponding to the
 $f^{\bm{0}}_{4,4}(\k)$.
(D) $B_{2g}$ form factor $\propto\sin(k_x)\sin(k_y)$
 driven by the attractive interactions (red arrows) A and B, and the repulsive
 interaction (blue arrow) C in the DW equation, where green lines denote nodes
 in the $B_{2g}$ form factor.
}
\label{fig:form-Cs122}
\end{figure}

Figure \ref{fig:form-Cs122}A shows $\q$ dependence of largest eigenvalue
$\lambda_{\q}$
with the self-energy for $r=0.96$ at $T=5$meV and that without the
self-energy for $r=0.30$ at $T=20$meV. $\lambda_{\q}$ becomes maximum at
$\q=\bm{0}$ and the dominant form factor
$f^{\bm{0}}_{4,4}(\k)\propto\sin(k_x)\sin(k_y)$ at $\q=\bm{0}$
is shown in Fig. \ref{fig:form-Cs122}B. As shown in Fig. \ref{fig:form-Cs122}C, this form factor corresponds to the
$B_{2g}$ next-nearest-neighbor bond order for $d_{xy}$ orbital, which is consistent with
experimentally observed $B_{2g}$ nematicity \cite{Cs122-B2g,RbFe2As2-nematic,Moroni-B2g,Shibauchi-Rb122}.
By analyzing the irreducible four-point vertex
$I^{\bm{0}}_{4,4;4,4}(k,k')$ in the DW Eq. (\ref{eqn:K}), we find that the
attractive (repulsive) interactions originate from the AL (MT) terms, as
shown in Fig. \ref{fig:form-Cs122}D. The obtained $\q=\bm{0}$
$B_{2g}$ bond order is derived from these interactions.
Since the AL terms are enhanced by the quantum interference between the spin
fluctuations with $\bm{Q}$ and $\bm{Q'}(=-\Q)$, as shown in Fig. \ref{fig:interference}D,
$\q=\bm{0}$ nematic bond order is realized.
The value of $\lambda_{\bm{0}}$ is strongly enhanced by the attractive
interactions for $d_{xy}$ orbital due to the AL terms. In this system, the nesting vector is
short $\Q\sim (0.5\pi,0)$, as shown in Fig. \ref{fig:form-Cs122}B. Due
to repulsive interaction by the MT terms, $f^{\bm{0}}_{4,4}(\k)$ changes
sign between $\k$ points on the FSs connected by $\Q$, as shown in
Fig. \ref{fig:form-Cs122}D. To summarize, the AL terms strongly enlarge
$\lambda_{\bm{0}}$ due to the paramagnon interference mechanism, and
the MT terms favor the $B_{2g}$ symmetry. Cooperation of the AL and MT terms
is important to realize the $B_{2g}$ bond order.

We comment on the recent experiments on RbFe$_2$As$_2$. The specific
heat jump at $T_S=40$K $(\Delta C/T_S)$ is very small
\cite{Shibauchi-Rb122}.  However, it is naturally understood based on
the recent theoretical scaling relation $\Delta C/T_S \propto T_S^b$
with $b\sim 3$ derived in Ref. \cite{Tazai-LW}.
Although the smallness of $B_{2g}$ nematic susceptibility in RbFe$_2$As$_2$ was
recently reported in Refs. \cite{Bohmer-B2g2,Bohmer-B2g}, the field-angle dependent specific
heat measurement has shown the finite $B_{2g}$ nematicity above $T_c$ \cite{Mizukami-B2g}. Further experimental and theoretical studies are necessary to
clarify the nematicity in AFe$_2$As$_2$ (A=Cs, Rb).

\begin{figure}[!htb]
\includegraphics[width=.6\linewidth]{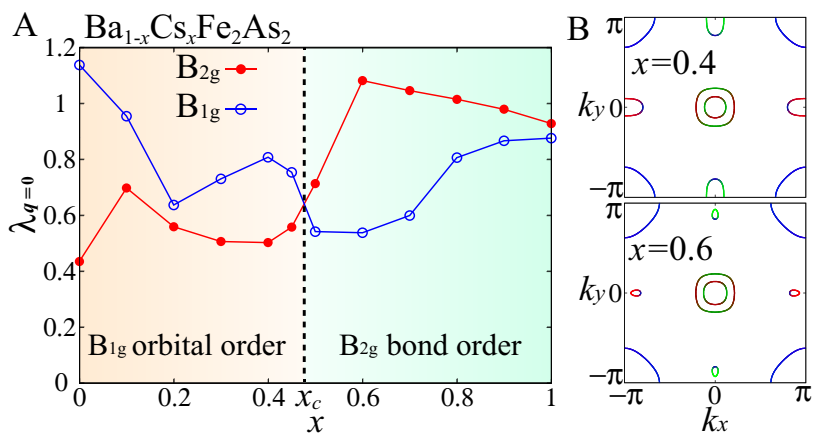}
\caption{
(A) $x$ dependences of $\lambda_{\q=\bm{0}}$ without the self-energy for $B_{1g}$ and $B_{2g}$
 symmetries in Ba$_{1-x}$Cs$_x$Fe$_2$As$_2$. 
(B) FSs for $x=0.4$ and $x=0.6$.
Dominant nematic order changes at $x=x_c\sim 0.5$ near the Lifshitz transition, where
 the electron FSs split into the four tiny Dirac pockets.
}
\label{fig:xdep-Cs122}
\end{figure}

Finally, we discuss $x$ dependence of nematicity in Ba$_{1-x}$A$_x$Fe$_2$As$_2$
(A=Cs, Rb). The schematic phase diagram in Ba$_{1-x}$Rb$_x$Fe$_2$As$_2$ given by the
experiment \cite{Shibauchi-Rb122} is shown in Fig. \ref{fig:interference}C. We introduce model Hamiltonian for
Ba$_{1-x}$Cs$_x$Fe$_2$As$_2$, by interpolating between 
BaFe$_2$As$_2$ model and CsFe$_2$As$_2$ model with the ratio $1-x:x$.
Figure \ref{fig:xdep-Cs122}A shows $x$ dependences of 
$\lambda_{\bm{q}=\bm{0}}$ without the self-energy for
the $B_{2g}$ and the $B_{1g}$ symmetries by fixing $T=30$meV and $r=0.30$.
Below $x=x_c\sim 0.5$, $B_{1g}$ nematic orbital order is dominant as
discussed in the previous section, while
$B_{2g}$ nematic bond order dominates over the $B_{1g}$ nematic orbital
order for $x>x_c$.
As shown in Fig. \ref{fig:xdep-Cs122}B, the Lifshitz transition occurs
at $x\sim x_c$, where the electron pockets split into the four tiny Dirac pockets.
Thus, the $B_{2g}$ nematic bond order appears when the nesting vector
$\Q$ between
the electron pockets and hole pocket around M point becomes short $\Q\sim(0.5\pi,0)$.
By taking account of the Lifshitz transition at $x\sim x_c$, the schematic phase diagram in Fig. \ref{fig:interference}C is also well
reproduced by the orbital/bond order due to the
paramagnon interference mechanism.
We note that the $\q=(0,\pi)$ smectic order is dominant over the
$\q=\bm{0}$ $B_{1g}$ nematic order at $x=0$, as shown in previous section.

\section{Conclusion}
\color{black}We discussed the rich variety of nematic/smectic states in Fe-based
superconductors in the same theoretical framework based on the
paramagnon interference mechanism. \color{black}
In this mechanism, the charge-channel order is induced
by the quantum interference between the spin fluctuations, as shown in
Fig. \ref{fig:interference}D. The form factor and wavevector of the DW
instability are derived from the DW equation based on the
paramagnon interference mechanism. Recently, a rigorous formalism of the DW
equation has been constructed based on the Luttinger--Ward (LW) theory in
Ref. \cite{Tazai-LW}. According to Ref.\cite{Tazai-LW}, the solution of
the DW equation gives the minimum of the grand potential in the LW theory. Thus, the
nematic/smectic order discussed in the present manuscript is
thermodynamically stable in the framework of the conserving approximation.

\begin{figure}[!htb]
\includegraphics[width=.8\linewidth]{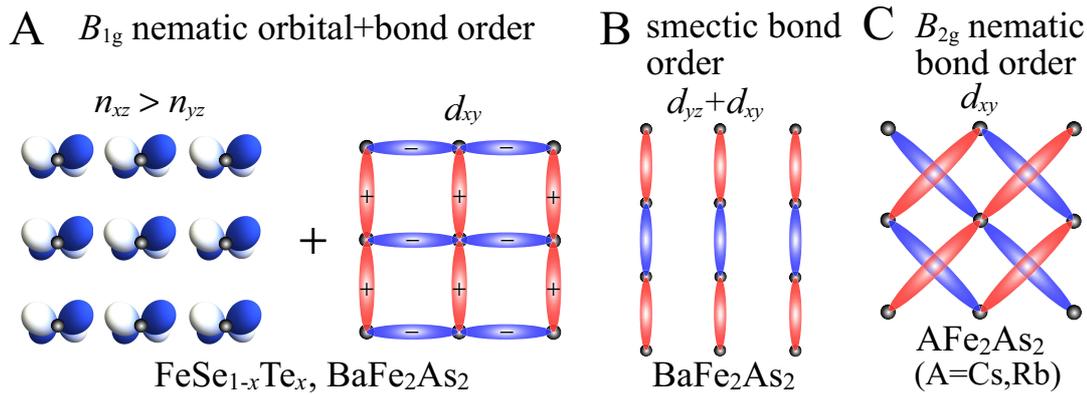}
\caption{
(A) Schematic picture of the $B_{1g}$ nematic orbital+bond order in
 FeSe$_{1-x}$Te$_x$ and BaFe$_2$As$_2$, where the orbital order for
 $d_{xz}$ and $d_{yz}$ orbitals coexists with the bond order for
 $d_{xy}$ orbital.
(B) Schematic picture of the smectic bond order for $d_{yz}$ and
 $d_{xy}$ orbitals in
 BaFe$_2$As$_2$.
(C) Schematic picture of the $B_{2g}$ nematic bond order for $d_{xy}$
 orbital in AFe$_2$As$_2$ (A=Cs, Rb).
}
\label{fig:order}
\end{figure}

By considering the characteristic
fermiology of each compound, the paramagnon interference
mechanism explains the rich variety of the nematic/smectic
states.
In Figs. \ref{fig:order}A, \ref{fig:order}B, \ref{fig:order}C, we
summarized the nematic/smectic orders revealed by the mechanism in the
present study. 
(i) In FeSe$_{1-x}$Te$_x$, each FS is very small and the $d_{xy}$-orbital
hole pocket is absent.
\color{black}In this case, the small spin fluctuations on the three orbitals cooperatively
lead to the $B_{1g}$ orbital order for $d_{xz}$ and
$d_{yz}$ orbitals coexisting with the $d_{xy}$-orbital bond order, as
shown in Fig. \ref{fig:order}A. \color{black} The
nematic orbital+bond order causes the Lifshitz transition, where the FS around Y point
disappears, consistently with the recent experiments.
The $x$ dependence of $T_S$ in the phase diagram is reproduced by
introducing the self-energy.
(ii) In BaFe$_2$As$_2$, the $d_{xy}$-orbital
hole pocket emerges. Since each electron and hole pocket is
relatively large and similar in size, the strong $d_{xy}$-orbital spin
fluctuations due to good nesting give rise to the 
smectic order shown in Fig. \ref{fig:order}B and the $B_{1g}$ nematic order. 
The smectic
order explains the tiny $T$-linear nematicity below $T=T^*(>T_S)$.
We predict the multistage transitions
with the smectic order at $T=T^*$ and nematic order at $T_S$.
(iii) In heavily hole-doped AFe$_2$As$_2$ (A=Cs, Rb),
 the tiny Dirac pockets around X(Y) point and the large $d_{xy}$-orbital hole
 pocket appear due to the hole-doping. The $B_{2g}$ bond order for
 the $d_{xy}$ orbital shown in Fig. \ref{fig:order}C emerges
 due to the $d_{xy}$-orbital paramagnon interference mechanism. The $B_{2g}$ bond order is triggered by the Lifshitz
transition of the electron FSs by the hole-doping.

\color{black}
The limitation of this theory is that the calculated VCs are reduced to the infinite series of the MT and AL terms. To verify the validity of the present theory, we performed the functional renormalization group (fRG) analysis for the single-orbital Hubbard model for cuprates \cite{Tsuchiizu-Cu} and the two-orbital Hubbard model for ruthenates \cite{fRG-Ru}, and obtained the bond-order (orbital order) in the former (latter) model. These results are consistent with experiments, and they are also obtained by the DW equation analysis. In the fRG theory, a huge number of higher-order VCs are generated in an unbiased way by solving the RG equation. Thus, the significance of the MT and AL terms in the present theory has been confirmed by the different and excellent theoretical framework.
 \color{black}

It is an important
future problem to clarify the mechanism of superconductivity and
non-Fermi-liquid behaviors of transport phenomena in the FeSe family by
considering the nematic fluctuations enlarged near the nematic QCP. This issue will be discussed in future publications \cite{Yamakawa-Bethe}.

\section*{Author Contributions}
S.O. performed all calculations with contributions from H.K.
S.O. and H.K wrote the paper.

\section*{Funding}
This work was supported
by Grants-in-Aid for Scientific Research from MEXT,
Japan (No. JP19H05825, No. JP18H01175, and No. JP17K05543), and
Nagoya University Research Fund.

\section*{Acknowledgments}
We acknowledge
 Y. Yamakawa, R. Tazai, and S. Matsubara for their collaboration in the theoretical
 studies. We are grateful to Y. Matsuda, T. Hanaguri, T. Shibauchi,
 S. Kasahara, T. Shimojima, and Y. Mizukami for useful discussions about
 experiments.

\bibliographystyle{frontiersinHLTH&FPHY} 
\bibliography{frontiers}
\end{document}